\documentclass[sigconf]{acmart}
\usepackage{tabularx}
\newcolumntype{Y}{>{\centering\arraybackslash}X}

\usepackage{algorithm}
\usepackage{algpseudocode}
\usepackage{multirow}
\usepackage{subcaption}
\usepackage{cleveref}
\algdef{S}[FOR]{ForEach}[1]{\algorithmicforeach\ #1\ \algorithmicdo}

\copyrightyear{2024} 
\acmYear{2024} 
\setcopyright{acmcopyright}
\acmConference[CIKM '24]{Proceedings of the 30st ACM International Conference on Information and Knowledge Management}{May 17--21, 2024}{Atlanta, GA, USA}
\acmBooktitle{Proceedings of the 30st ACM International Conference on Information and Knowledge Management (CIKM '24), May 17--21, 2024, Atlanta, GA, USA}
\acmPrice{15.00}
\acmDOI{10.1145/3511808.3557065}
\acmISBN{978-1-4503-9236-5/25/10}
\begin{document}

\title{EnhancedRL: An Enhanced-State Reinforcement Learning Algorithm for Multi-Task Fusion in Recommender Systems}

\author{Peng Liu}
\affiliation{
    \institution{Tencent Inc.}
    \city{Beijing}
    \country{China}
}
\email{liupengvswww@gmail.com}
\orcid{0009-0000-7271-4721}

\author{Cong Xu}
\affiliation{
  \institution{Tencent Inc.}
  \city{Beijing}
  \country{China}
}
\email{congcxu@tencent.com}

\author{Jiawei Zhu}
\affiliation{
    \institution{Tencent Inc.}
    \city{Beijing}
    \country{China}
}
\email{erickjwzhu@tencent.com}

\author{Ming Zhao}
\affiliation{
  \institution{Tencent Inc.}
  \city{Beijing}
  \country{China}
}
\email{marcozhao@tencent.com}

\author{Bin Wang}
\affiliation{
  \institution{Tencent Inc.}
  \city{Beijing}
  \country{China}
}
\email{hillmwang@tencent.com}

\renewcommand{\shortauthors}{Peng Liu et al.}

\begin{abstract}
As a key stage of Recommender Systems (RSs), Multi-Task Fusion (MTF) is responsible for merging multiple scores 
output by Multi-Task Learning (MTL) into a single score, finally determining the recommendation results. 
Recently, Reinforcement Learning (RL) has been applied to MTF to maximize long-term user satisfaction within a recommendation session. 
However, due to limitations in modeling paradigm, all existing RL algorithms for MTF can only utilize user features and statistical features 
as the state to generate actions at the user level, but unable to leverage item features and other valuable features, 
which leads to suboptimal performance. Overcoming this problem requires a breakthrough in the existing modeling paradigm, 
yet, to date, no prior work has addressed it.

To tackle this challenge, we propose EnhancedRL, an innovative RL algorithm. 
Unlike existing RL-MTF methods, EnhancedRL takes the enhanced state as input, 
incorporating not only user features but also item features and other valuable information. 
Furthermore, it introduces a tailored actor-critic framework - including redesigned actor and critics and a novel learning procedure - 
to optimize long-term rewards at the user-item pair level within a recommendation session. 
Extensive offline and online experiments are conducted in an industrial RS and 
the results demonstrate that EnhancedRL outperforms other methods remarkably, 
achieving a $+3.84\%$ increase in user valid consumption and a $+0.58\%$ increase in user duration time. 
To the best of our knowledge, EnhancedRL is the first work to address this challenge, 
and it has been fully deployed in a large-scale RS since September 14, 2023, yielding significant improvements.
\end{abstract}



\maketitle

%%%%%%%%%%%%%%%%%%%%%%%%%%%%%%%%%%%%%%%%%%%%%%%%%%%%%%%%%%%%%%%%%%%%%%%%%%%%%%%%%%%%%%
%%%%%%%%%%%%%%%%%%%%%%%%%%%%%%%%   Introduction   %%%%%%%%%%%%%%%%%%%%%%%%%%%%%%%%%%%%
%%%%%%%%%%%%%%%%%%%%%%%%%%%%%%%%%%%%%%%%%%%%%%%%%%%%%%%%%%%%%%%%%%%%%%%%%%%%%%%%%%%%%%
\section{Introduction}
\label{sec:intro}
Recommender Systems (RSs) \cite{ref1, ref2} provide personalized recommendation service based on user interest, 
which are widely used in various platforms such as short video platforms \cite{ref3, ref7, ref14}, E-commerce platforms \cite{ref6, ref8, ref9, ref10, ref11}, 
video platforms \cite{ref4, ref5} and social networks \cite{ref12, ref13, ref40}, serving billions of users every day. 
In short, industrial RSs generally consist of three main stages: candidate generation, ranking, and multi-task fusion (MTF) \cite{ref4, ref15, ref41}.
During candidate generation stage, thousands of candidates are selected from millions or even more items.
The ranking stage typically uses a Multi-Task Learning (MTL) model \cite{ref4, ref8, ref16, ref17, ref18, ref42} to predict scores for various user behaviors, 
such as valid click, watching time, fast slide, like, and sharing. 
Finally, an MTF model combines the multiple scores output by the MTL model into a final score \cite{ref15, ref41} 
and thus determines the ultimate recommendation results.

\begin{figure}[hbtp!]
  \centering
  \includegraphics[width=0.9\linewidth]{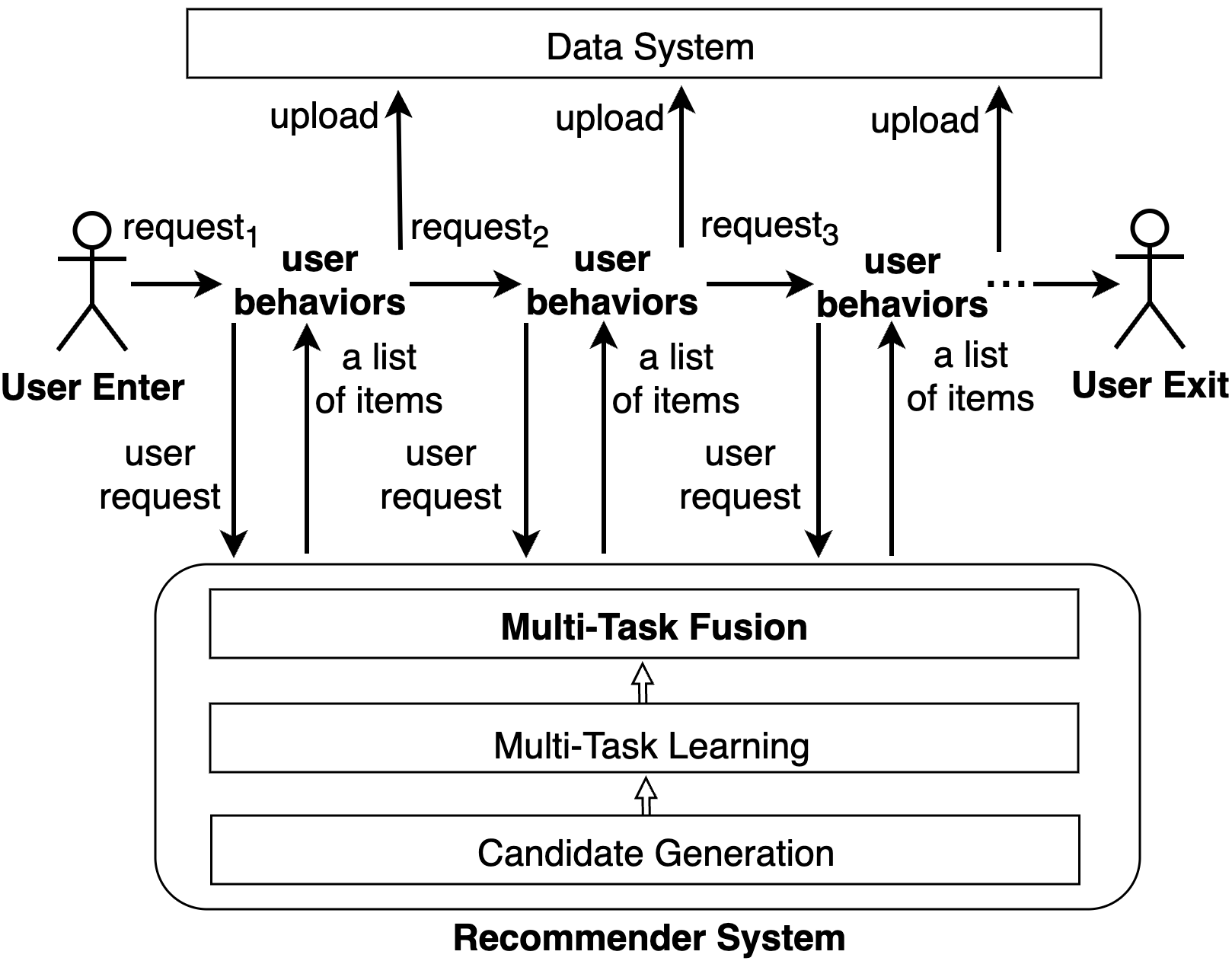}
  \caption{The interactions between a user and an RS within a recommendation session.}
  \label{fig:session}
\end{figure}

The objective of MTF in RSs is to optimize user satisfaction, typically measured by the weighted sum of various types of user feedback 
within a single recommendation or a recommendation session. 
A recommendation session refers to the period from when a user first accesses the RS until the user exits, 
and may consist of one or more consecutive interactions, as shown in Figure \ref{fig:session}. 
Early works, such as Grid Search \cite{ref36} and Bayesian Optimization \cite{ref19}, select optimal fusion weights by searching parameter combinations, 
which are inefficient and assign identical fusion weights to all users. 
Evolution Strategy (ES) \cite{ref20, ref21, ref22} takes only a small number of features as input to generate personalized fusion weights.
However, due to its learning pattern, the capability of ES is limited \cite{ref23}. 
Furthermore, all the methods mentioned above merely focus on the reward of current recommendation 
(also called instant reward or instant user satisfaction), but overlooking long-term rewards.

In personalized RSs, such as YouTube, Shorts, TikTok and Kwai, the current recommendation significantly influences subsequent recommendations 
within the same recommendation session. Therefore, it is essential to jointly consider both the instant reward of the current recommendation 
and the potential rewards of future recommendations, which is well-suited to be solved using Reinforcement Learning (RL) \cite{ref26}. 
A few studies \cite{ref15, ref24, ref25, ref41} have used RL algorithms for MTF to maximize long-term rewards. 
However, existing RL-MTF methods are constrained by their modeling paradigm, as their states can only utilize user and statistical features, 
and generate actions solely at user level, which fails to incorporate item features and other valuable features, leading to suboptimal outcomes. 
Although RL has been applied to various problems in RSs, this paper focuses specifically on RL algorithms for MTF to optimize user satisfaction. 
Other topics, such as learnable fusion formulas \cite{ref44} or user-retention optimization \cite{ref45}, are beyond the scope of this study.

Addressing this limitation requires a fundamental shift in the RL-MTF modeling paradigm: the state, reward, and the learning processes of the actor and critic 
must be redefined. To solve this challenge, we propose an innovative algorithm called EnhancedRL. 
Unlike existing methods, EnhancedRL takes the enhanced state as input, incorporating not only user features but also item features and other valuable information. 
Furthermore, it adopts a new modeling paradigm with redesigned actor, critic, and a learning process tailored for MTF, 
aiming to optimize long-term rewards at the user-item pair level. 

In this paper, our contributions can be summarized as follows:
\begin{itemize}
\item We point out that the existing RL-MTF modeling paradigm is not fully aligned with the MTF requirements in the context of RSs.
It restricts state features to user features and statistical features and generates fusion weights merely at user level, 
but can't utilize item features and other valuable information, resulting in suboptimal performance. 

\item We propose EnhancedRL, which adopts an innovative modeling paradigm to address this challenge. 
Unlike existing methods, EnhancedRL puts forward the concept of hierarchical state that includes not only user features but also item features and other relevant information. 
It then introduces a novel actor, a novel critic, and a new learning procedure, all designed to optimize long-term rewards at the user-item pair level.

\item Extensive offline and online experiments are conducted in an industrial RS and the results demonstrate that EnhancedRL outperforms 
other methods remarkably, achieving a $+3.84\%$ increase in user valid consumption and a $+0.58\%$ increase in user duration time. 

\item To the best of our knowledge, EnhancedRL is the first method designed to maximize long-term user satisfaction at the user-item pair level 
within a recommendation session, and it has been deployed in a large-scale RS.
\end{itemize}

We possess substantial know-how—practical, hands-on experience and deep technical mastery of key RL-MTF aspects such as algorithm optimization, 
reward design, efficient exploration, and training stability.
During our spare time at work, we intend to compile the findings of these work efforts into academic papers for public publication.

%%%%%%%%%%%%%%%%%%%%%%%%%%%%%%%%%%%%%%%%%%%%%%%%%%%%%%%%%%%%%%%%%%%%%%%%%%%%%%%%%%%%%%
%%%%%%%%%%%%%%%%%%%%%%%%%%%%%%%%   Related Work   %%%%%%%%%%%%%%%%%%%%%%%%%%%%%%%%%%%%
%%%%%%%%%%%%%%%%%%%%%%%%%%%%%%%%%%%%%%%%%%%%%%%%%%%%%%%%%%%%%%%%%%%%%%%%%%%%%%%%%%%%%%
\section{Related Work}
\label{related_work}
In RSs, Multi-Task Fusion (MTF) \cite{ref4, ref15, ref41} merges multiple scores predicted by MTL 
\cite{ref4, ref8, ref16, ref17, ref18, ref42} into a single score, 
which determines the final ranking of candidates. At first, Grid Search \cite{ref36} is applied to find the optimal fusion weights 
by searching through a candidate set filled with lots of parameter combinations. 
Following this, Bayesian Optimization \cite{ref19, ref37} is introduced to accelerate the parameter search process. 
However, these two methods are inefficient and merely generate the same fusion weights for all users. 
ES \cite{ref20, ref21, ref22} considers user preference and generate personalized fusion weights for different users. 
But the learning pattern of ES, which is based on perturbation and selection, limits its performance \cite{ref23}. 
Moreover, all the above methods merely focus on instant rewards but neglecting long-term rewards. 

\begin{figure}[hbtp!]
  \centering
  \includegraphics[width=0.86\linewidth]{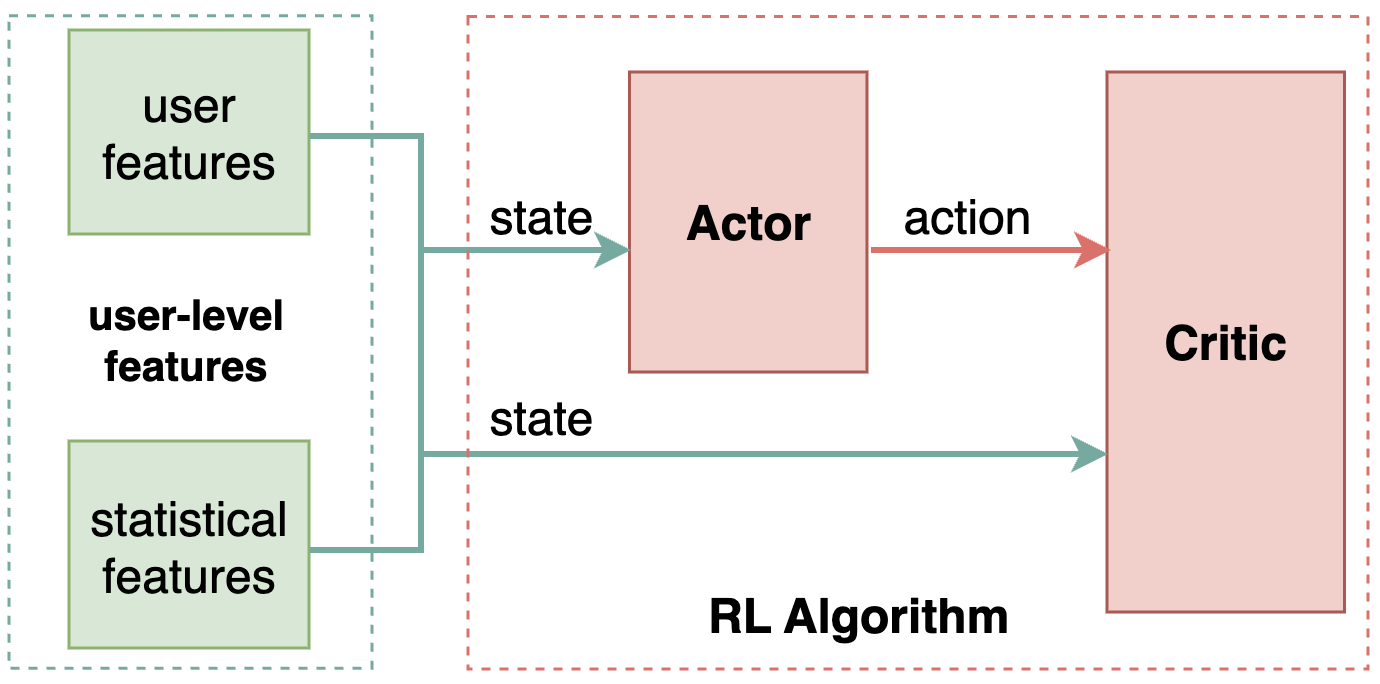}
  \caption{The modeling paradigm of existing RL-MTF can only use user level features as the state and generates actions at user level within a recommendation session.}
  \label{fig:old_pattern}
\end{figure}
In personalized RSs \cite{ref3, ref7, ref14}, current recommendation has an obvious impact on subsequent recommendations within a session. 
Therefore, a few studies have utilized RL to optimize long-term rewards. 
For instance, \cite{ref38} employs RL to search the optimal weights between the predicted click-through rate and the advertiser's bid price. 
\cite{ref15} proposes BatchRL-MTF to search the optimal fusion weights, which has been deployed in multiple RSs in Tencent for several years. 
Furthermore, other teams have utilized alternative RL algorithms for MTF, including DDPG \cite{ref27}, CQL+SAC \cite{ref30, ref39}, and IQL \cite{ref31}, among others. 
However, these RL-MTF methods have the following drawbacks: 
first, their constraints are too strict in order to avoid Out-of-Distribution (OOD), which significantly hurt their performance; 
second, they are unaware of the exploration policy; 
third, the efficiency of their exploration policies \cite{ref15, ref32} are inefficient. 
Therefore, UnifiedRL \cite{ref41} is proposed to solve these problems and has achieved remarkable improvements in several industrial RSs. 
However, due to their modeling paradigm, existing RL-MTF methods can use only user and statistical features as the state to generate actions (fusion weights) 
at the user level within a recommendation session, but are unable to harness item features and other valuable features (user-item pair level features), 
which leads to suboptimal performance, as shown in Figure \ref{fig:old_pattern}.

%%%%%%%%%%%%%%%%%%%%%%%%%%%%%%%%%%%%%%%%%%%%%%%%%%%%%%%%%%%%%%%%%%%%%%%%%%%%%%%%%%%%%%
%%%%%%%%%%%%%%%%%%%%%%%%%%%%%%  Problem Definition  %%%%%%%%%%%%%%%%%%%%%%%%%%%%%%%%%%
%%%%%%%%%%%%%%%%%%%%%%%%%%%%%%%%%%%%%%%%%%%%%%%%%%%%%%%%%%%%%%%%%%%%%%%%%%%%%%%%%%%%%%
\section{Problem Definition}
\label{sec:problem_defination}
In this section, we present the problem definition of EnhancedRL, which is designed to optimize cumulative reward within a recommendation session 
at the user-item granularity and differs from existing RL-MTF modeling paradigms.

As mentioned earlier, the current recommendation clearly influences subsequent recommendations within the same session. 
At each time step $t$, after an RS receives a user's request, it performs the following steps. 
First, thousands of candidate items are selected from millions (or more) of items. 
Second, the MTL model predicts scores for multiple user behaviors for each candidate item. 
Third, the MTF model generates fusion weights to combine the scores produced for each item into a final score using Eq. \ref{eq:merge_eq}. 
This fusion formula is commonly used in RSs, where $pred\_score_i$ ($i \in [1, k]$) denotes the score produced by the MTL model, 
and $power_i$ and $bias_i$ are elements of the action which is a vector composed of floating-point numbers output by RL model. 
Finally, a list of items is sent to the user, and the user's feedback is reported to the RS's data system.
\begin{equation}
\label{eq:merge_eq}
    final\_score = \prod\limits_{i=1}^{k}\left(pred\_score_{i} + bias_{i}\right)^{power_{i}} \,.
\end{equation}

It should be noted that \textbf{EnhancedRL generates a separate action for each candidate item, 
a significant departure from the prevailing RL-MTF modeling paradigm, which produces a single action for all candidates}. 
Consequently, EnhancedRL's state and reward definitions, along with its actor and critic, 
differ substantially from those in existing RL-MTF methods.

We model the above fusion problem within a recommendation session as a Markov Decision Process (MDP). 
In this MDP, RS acts as an agent that interacts with a user (environment) and makes sequential recommendations, 
with the goal of maximizing the cumulative reward within a session.
The MDP framework has the following essential components \cite{ref15, ref26, ref27, ref28, ref29, ref41}:

• \textbf{State Space ($\mathcal{S}$):} \textbf{$\mathcal{S}$} is a set of state $s$ which includes not only user features 
(e.g., age, gender, top k interests, flush num, user behavior sequence), 
but also item features and other valuable features (e.g., predictions output by MTL and other features, etc.). 
This differs from existing RL-MTF methods, which mainly use user-level features as the state. 
To emphasize this distinction, we call it the enhanced state. For convenience, we will also refer to it simply as the ``state''.

• \textbf{Action Space ($\mathcal{A}$):} \textbf{$\mathcal{A}$} is a set of action $a$ generated by RL model. 
In the context of MTF, action $a$ is a fusion weight vector 
($a_{1}$, . . ., $a_{k}$), of which each element corresponds to different power or bias term in Eq. \ref{eq:merge_eq}. 
As mentioned above, in EnhancedRL, each action corresponds to a user-item pair.

• \textbf{Reward ($\mathcal{R}$):} After RS takes an action $a_{tj}$ at state $s_{tj}$ for each user-item pair and sends a list of items to a user, 
the user's various behaviors to those items will be reported to data system. 
The instant reward $r(s_{tj}, a_{tj})$ for each user-item pair will be calculated based on these behaviors, as shown in Eq. \ref{eq:pair_reward}. 
The instant reward for current recommendation $r(s_t, a_t)$ is the sum of instant rewards for all user-item pairs, as shown in Eq. \ref{eq:current_reward}

• \textbf{Discount Factor ($\gamma$):} \textbf{$\gamma$} determines how much weight the agent assigns to future rewards 
compared to instant reward for current recommendation, $\gamma$ $\in$ $[0, 1]$.

With the definitions above, applying RL to MTF can be summarized as follows: 
given the interaction history between the RS and a user, represented as an MDP within a session, 
the goal is to learn the optimal policy to maximize cumulative reward.

%%%%%%%%%%%%%%%%%%%%%%%%%%%%%%%%%%%%%%%%%%%%%%%%%%%%%%%%%%%%%%%%%%%%%%%%%%%%%%%%%%%%%%
%%%%%%%%%%%%%%%%%%%%%%%%%%%%%%%%       Method     %%%%%%%%%%%%%%%%%%%%%%%%%%%%%%%%%%%%
%%%%%%%%%%%%%%%%%%%%%%%%%%%%%%%%%%%%%%%%%%%%%%%%%%%%%%%%%%%%%%%%%%%%%%%%%%%%%%%%%%%%%%
\section{Method}
\label{sec:method}
\subsection{Reward Function}
\label{sec:reward_fun}
To evaluate user satisfaction at different levels, we define corresponding rewards. 
First, to evaluate the user's satisfaction with the specific item, the instant reward for the user-item pair is defined as shown in Eq. \ref{eq:pair_reward}, 
where $w_i$ is the weight of behavior $\upsilon_{i}$. In our recommendation scenario, user behaviors ($\upsilon_{1}$, ..., $\upsilon_{k}$) 
contains watching time, valid consumption (watching a video longer than $10$ seconds) and interaction behaviors such as like, sharing, collecting, etc. 
By analyzing the correlations of different user behaviors with the long-term optimization goal, we assign different weights to the behaviors.
Note that long-term optimization objective may vary across different recommendation scenarios and can be adjusted over time. 
Second, to evaluate the user's satisfaction with current recommendation, the instant reward $r(s_t, a_t)$ is the sum of instant reward for the user-item pairs 
included in this recommendation, as shown in Eq. \ref{eq:current_reward}, where $l$ is the length of the recommendation list.
\begin{equation}
\label{eq:pair_reward}
    r \left(s_{tj}, a_{tj}\right) = \sum\limits_{i = 1}^{k}w_{i} * \upsilon _{i} \,. 
\end{equation}

\begin{equation}
\label{eq:current_reward}
    r \left(s_{t}, a_{t}\right) = \sum\limits_{j = 1}^{l}  r(s_{tj}, a_{tj}) \,. 
\end{equation}

The calculation of cumulative reward within a session, denoted as $G_t$, is presented in Eq. \ref{eq:cul_reward}. 
$G_t$ represents the cumulative reward starting from time step $t$. 
$r_{t+i}(s_{t+i}, a_{t+i})$ denotes the instant reward for the recommendation received at step $t + i$. 
$\gamma$ is the discount factor and $T$ indicates the terminal time step.

\begin{equation}
\label{eq:cul_reward}
  G_t = \sum_{i=0}^{T-t} \gamma^i * r_{t+i}(s_{t+i}, a_{t+i}) \,. 
\end{equation}

\subsection{Online Exploration}
\label{sec:online_exploration}
Before training, a large amount of exploration data must be collected, which critically impacts model performance. 
EnhancedRL adopts the exploration approach proposed by \cite{ref41}, which mitigates the inefficiency and adverse effects on user experience 
found in most online exploration methods for RL-MTF \cite{ref15, ref26, ref32}. 
UnifiedRL notices that the action generated by the new learned RL policy (actor) will typically not deviate far from the action generated by baseline RL policy 
for the same state. Inspired by this finding, a simple but extremely efficient exploration policy is designed \cite{ref41}, 
which defines personalized exploration upper and lower bounds based on baseline policy, as shown in Eq. \ref{eq:explore_policy}. 
Exploration action is generated by the action output by baseline policy $\mu_{bp}$ plus a random perturbation $\varepsilon$ 
generated by uniform distribution defined by $b_l$ and $b_u$, which should be selected carefully.
\begin{equation}
\label{eq:explore_policy}
    \mu _{ep}(s) = \mu_{bp}(s) + \varepsilon , \enspace \varepsilon \sim U(b_{l}, b_{u}) \,.
\end{equation}

\begin{figure}[hbtp!]
  \centering
  \includegraphics[width=0.95\linewidth]{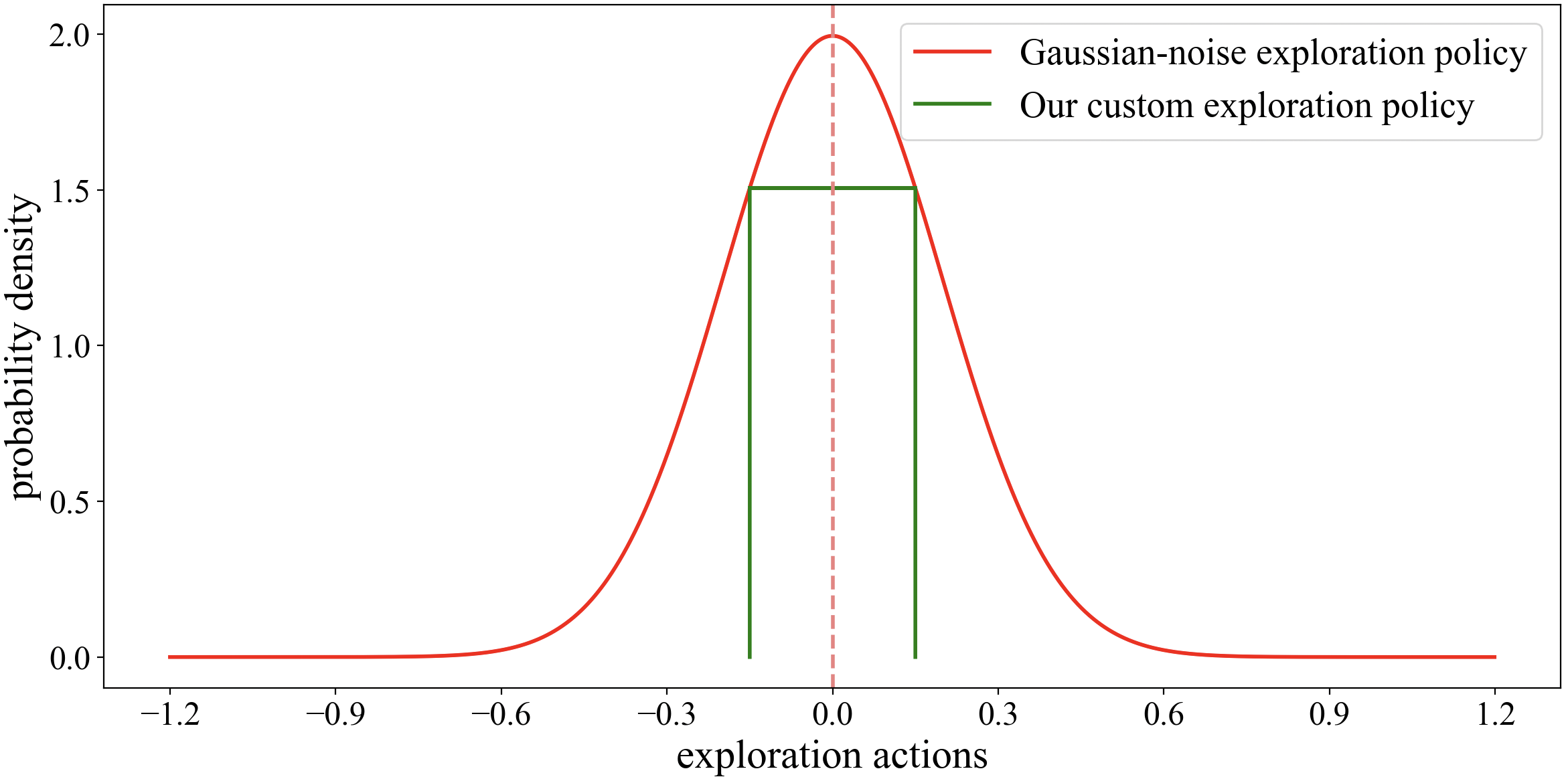}
  \caption{Comparison between the action distribution of our custom exploration policy and that of the Gaussian-noise exploration policy. 
  }
  \label{fig:explore_dist}
\end{figure}

The basic idea of this exploration policy is to eliminate low-value exploration space and only focus on exploring potential high-value areas, 
as shown in Figure \ref{fig:explore_dist}. 
Under the experiment setup of \cite{ref41}, it is roughly $2^{10}$ times more efficient than the commonly used Gaussian-noise exploration policy. 
As the number of dimensions of the action increases, efficient exploration becomes more valuable. 
Furthermore, the progressive training mode presented in \cite{ref41} is also adopted, allowing for smaller upper and lower bounds to be set for personalized exploration space.

\subsection{EnhancedRL}
\label{sec:method_details}
\begin{figure}[hbtp!]
  \centering
  \includegraphics[width=1.05\linewidth]{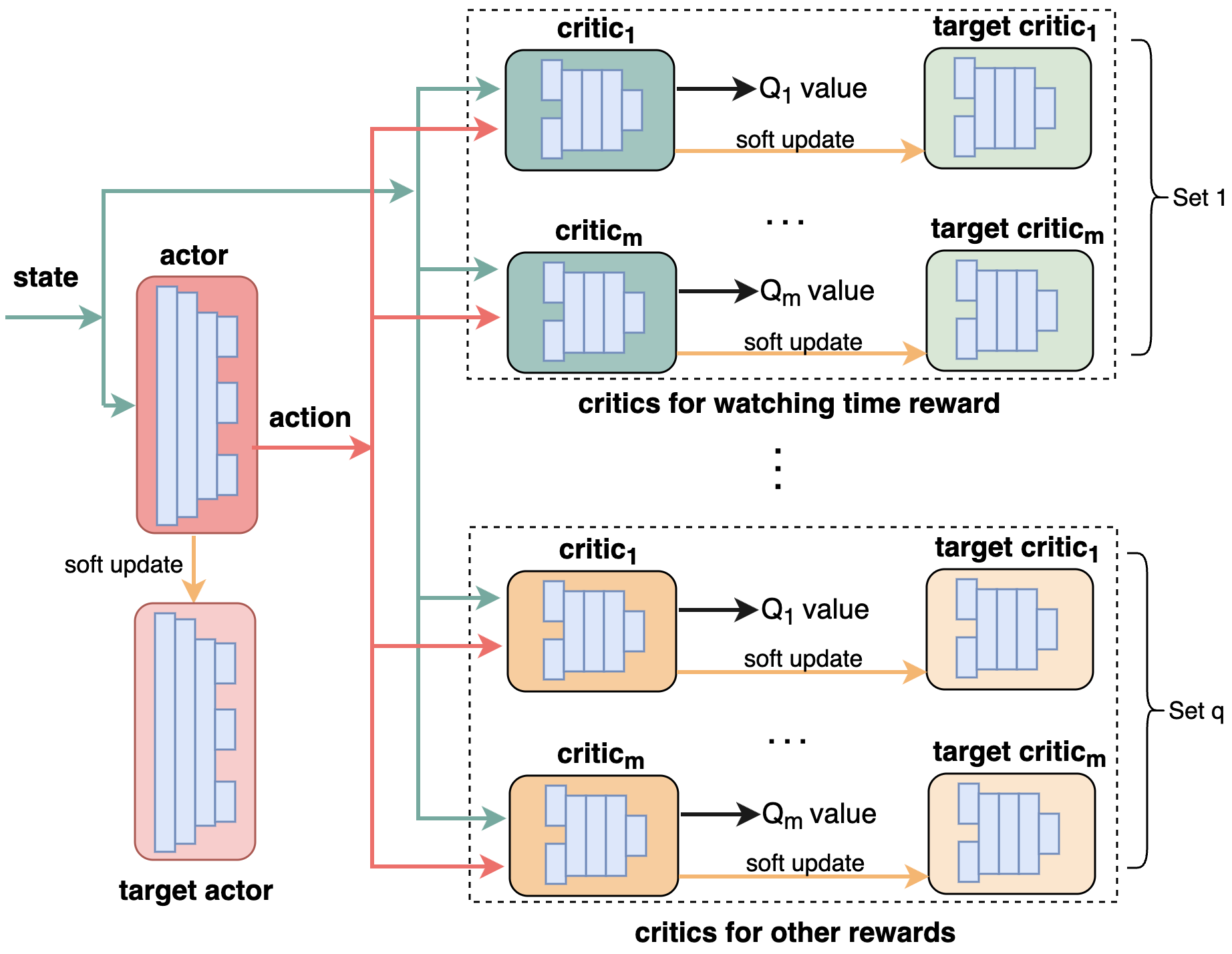}
  \caption{The framework of EnhancedRL, which consists of an actor and $q$ sets of critics, along with corresponding target actor and target critics.}
  \label{fig:enhancedRL}
\end{figure}

The framework of EnhancedRL is shown in Figure \ref{fig:enhancedRL}. The key components of EnhancedRL will be introduced as below.

\begin{figure}[hbtp!]
  \centering
  \includegraphics[width=1.0\linewidth]{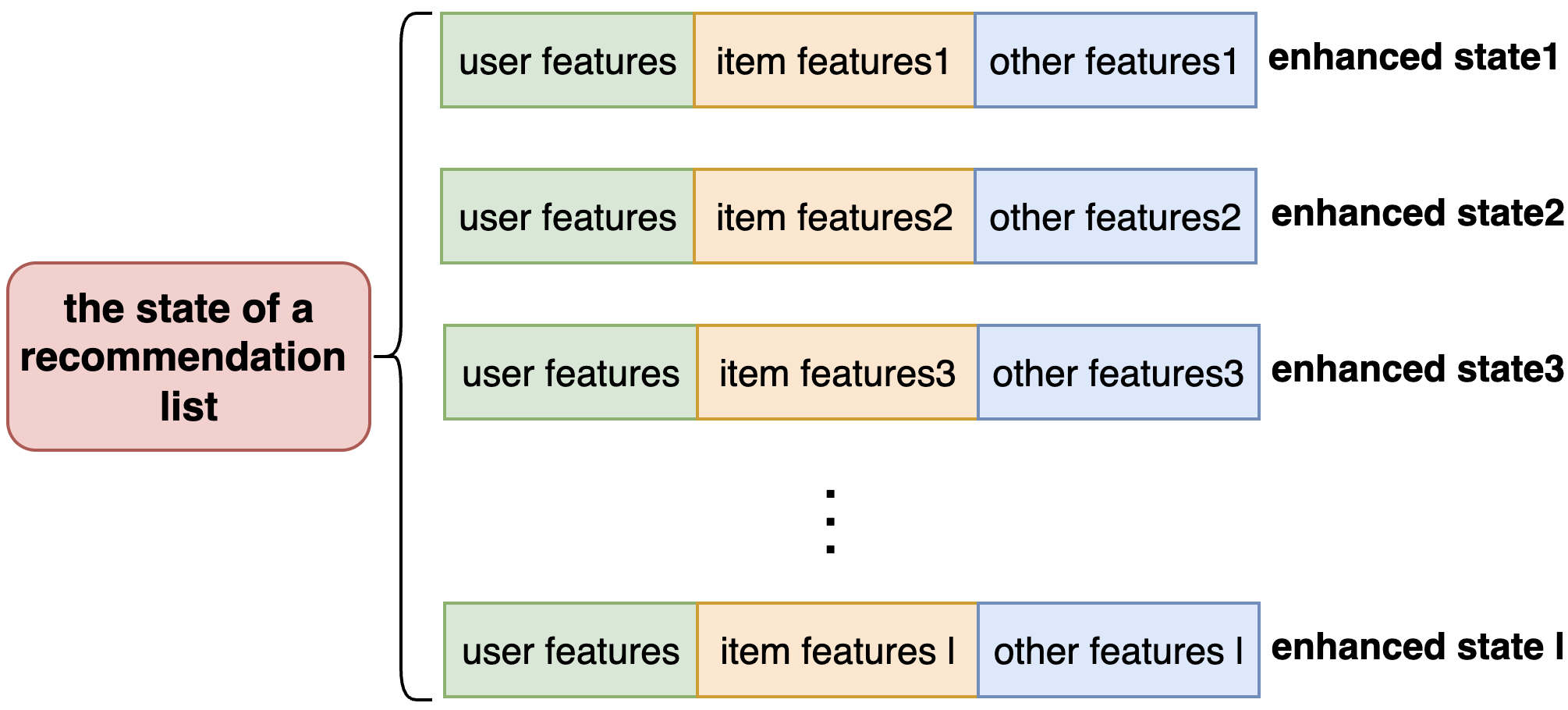}
  \caption{Hierarchical state in EnhancedRL: an enhanced state comprises user features, item features, and other contextual features; 
  the state of a recommendation list consists of $l$ enhanced states.}
  \label{fig:state_layer}
\end{figure}

\subsubsection{Actor}
\label{sec:sec_actor}
The goal of the actor $\mu(s)$ is to generate the optimal actions for given states to maximize long-term reward within a recommendation session. 
The actor's structure is similar to that of PLE \cite{ref18}. As it is not a core part of EnhancedRL, it will not be described in detail. 
As mentioned before, the state of EnhancedRL includes not only user features but also item features and other valuable features, corresponding to a user-item pair, 
which is different from existing RL-MTF methods. Therefore, the EnhancedRL's actor is redefined based on the enhanced states (user-item pair granularity) to 
produce the optimal actions to maximize long-term reward in a session, as shown in Eq. \ref{eq:actor1} and Eq. \ref{eq:actor2}.

The actor in EnhancedRL smoothly integrates the offline model with its online exploration policy. 
The first term of Eq. \ref{eq:actor1} represents the negative of the weighted sum of the means which are computed using the values estimated 
by the corresponding sets of critics, described in detail in Section \ref{sec:sec_critic}. 
The weight $w_i$ here is the same as it in Eq. \ref{eq:pair_reward}. 
$q$ denotes the number of critic sets, and $m$ denotes the number of critics in each group. 
During the training of $\mu(s)$, the upper and lower bounds of the exploration data distribution for each enhanced state can be directly obtained, 
as mentioned in Section \ref{sec:online_exploration}, which is leveraged to relax overly strict constraints and fully exploit the capacity of $\mu(s)$. 
If the action generated by $\mu(s)$ at state $s_{tj}$ falls within its upper and lower bounds, 
the value of the second term in Eq. \ref{eq:actor1} is zero—meaning no penalty is imposed, thereby avoiding harm to the model's performance.
Otherwise, a penalty is applied based on the extent to which the action exceeds either the upper or lower bound.
In this way, the performance of the actor is significantly improved compared to existing RL-MTF methods, 
as demonstrated by the experiments in Section \ref{sec:experiments}. 
\begin{align}
\label{eq:actor1}
  &\theta^{k+1} \longleftarrow arg\min\limits_{\theta^{k} }E_{s_{tj}\sim \mathcal{D},a\sim{\mu(s_{tj}|\theta^{k})}} \;
     \Bigg[- \sum_{i=1}^q w_i \bar{Q}_i \nonumber \\
      & \qquad + \eta  d(\mu(s_{tj}|\theta^{k})) + \lambda \sum_{i=1}^q w_i \left(
        \left(
          \frac{1}{m} \sum_{b=1}^m \left( Q_{ib} - \bar{Q_i} \right)^2
        \right)^{1/2}
      \right) 
    \Bigg] \,. \\ 
  & \qquad \mu_{tj}^k = \mu(s_{tj} \mid \theta^k); \ Q_{ib} = Q_{ib}(s_{tj},\, \mu_{tj}^k); \ \bar{Q}_i = \frac{1}{m} \sum_{b=1}^m Q_{ib} \nonumber 
\end{align}

\begin{align}
\label{eq:actor2}
    d(\mu(s_{tj}|\theta^{k})) = \left\{
      \begin{array}{l} 
      0, \quad \ \ if \ \mu(s_{tj}|\theta^{k}) \in [\mu_{bp}(s_{tj}) + b_{l}, \mu_{bp}(s_{tj}) + b_{u}] \\ \\
      \omega  \exp \left(\displaystyle\frac {\mu(s_{tj}|\theta^{k}) - (\mu_{bp}(s_{tj}) + b_{u})} 
      {\beta  (b_{u} - b_{l})} \right),  \qquad \qquad  \\
      \qquad if \ \mu(s_{tj}|\theta^{k}) > \mu_{bp}(s_{tj}) + b_{u} \\ 
      \\ \omega  \exp \left(\displaystyle\frac{(\mu_{bp}(s_{tj}) + b_{l}) - \mu(s_{tj}|\theta^{k})} 
      {\beta  (b_{u} - b_{l})} \right), \\
      \qquad if \ \mu(s_{tj}|\theta^{k}) < \mu_{bp}(s_{tj}) + b_{l}
    \end{array}
    \right.
\end{align}

Furthermore, a penalty mechanism is introduced as the third term, defined as the weighted sum of the standard deviations of the estimated values 
produced by the corresponding sets of critics \cite{ref33} to alleviate potential data distribution imbalance. 
Owing to the high efficiency of the custom exploration policy, the exploration actions have a significantly higher average density 
compared to those generated by Gaussian-noise exploration policy, which is highly beneficial for model performance. 
Moreover, compared with Gaussian perturbation, random perturbation within personalized upper and lower bounds reduces 
the impact of imbalanced data distribution on model training. 
It is important to emphasize that EnhancedRL generates an action for each user-item pair, 
rather than just generating a single action for all candidates like existing works \cite{ref15, ref24, ref25, ref41}. 

The target actor $\mu^\prime(s)$ is an auxiliary model responsible for generating the next optimal action for the next state 
to alleviate the overestimation problem caused by bootstrapping, whose parameters are periodically soft updated using the actor model $\mu(s)$.
Other approaches to mitigate overestimation can also be used \cite{ref26, ref32}.

\subsubsection{Critic}
\label{sec:sec_critic}
The critic $Q(s, a)$ is in charge of estimating the reward for a given state-action pair $(s_{tj}, a_{tj})$ within a recommendation session. 
$Q(s, a)$ also integrates critic network with its exploration policy to avoid OOD problem, as shown in Eq. \ref{eq:critic1} and Eq. \ref{eq:critic2}. 
To maximize the cumulative reward within a session at the granularity of user-item pairs, 
a novel mechanism is presented that aggregates the rewards of the user-item pairs in current recommendation list into the current recommendation reward, 
as shown in Figure \ref{fig:state_layer}, and uses adjacent cumulative rewards to compute temporal-difference (TD) updates. 

In EnhancedRL, each key reward is associated with a set of critics: each set comprises $m$ independent critics, and there are $q$ sets in total. 
Less important rewards are aggregated into the key rewards. The choice of these hyperparameters ($m$ and $q$) is determined by 
available resource and reflects a trade-off between cost and performance. 
In our scenario, we set $m$ = $10$ and $q$ = $2$. In addition, each recommendation list for a user request contains $l$ = $5$ videos.
One set of critics is responsible for estimating the cumulative reward associated with watching time, 
while the other set focuses on the cumulative rewards of other user behaviors. All the critics are initialized and trained independently. 

To alleviate overestimation, we define a target critic $Q^\prime(s, a|\phi^\prime)$ for each critic $Q(s, a|\phi)$, 
whose parameters are periodically soft-updated using $Q(s, a|\phi)$. 
If the next action generated by $\mu^\prime(s)$ at next state $s_{(t+1)j}$ falls within the exploration's upper and lower bounds, 
the value of $Q^\prime(s_{(t+1)j}, \mu^\prime(s_{(t+1)j})|\phi^\prime)$ in Eq. \ref{eq:critic1} will not be punished. 
Otherwise, a penalty is applied based on the deviation that surpasses the upper or lower bounds of the exploration, as shown in Eq. \ref{eq:critic2}. 

\begin{align}
\label{eq:critic1}
  \phi^{k+1} \longleftarrow	 \  &arg\min\limits_{\phi^{k}} 
  E_{s_{tj}\sim \mathcal{D}, a_{tj} \sim \mathcal{D}, s_{(t+1)j} \sim \mathcal{D} ,a_{(t+1)j} \sim \mu^\prime(s_{(t+1)j})} \biggl[ \quad \nonumber \\
  &\bigg( \sum_{j=1}^{l}  Q(s_{tj}, a_{tj}|\phi^{k}) \ - \ \Big(\sum_{j=1}^{l}r(s_{tj}, a_{tj}) \ + \nonumber \\
  &\gamma  \varphi(\mu^\prime(s_{(t+1)j}))  Q^\prime(s_{(t+1)j}, \mu^\prime(s_{(t+1)j})|\phi^\prime) \Big) \bigg)^{2} \biggr]  \\
  & for \ j = 1, ..., l. \nonumber
\end{align}

\begin{align}
\label{eq:critic2}
  \varphi(\mu^\prime(s_{(t+1)j})) = \left\{
    \begin{array}{l} 
    1, \  if \ \mu^\prime(s_{(t+1)j}) \in [\mu_{bp}(s_{(t+1)j}) + b_{l}, \\ \qquad \qquad \qquad \qquad \mu_{bp}(s_{(t+1)j}) + b_{u}] \\ \\
    \varpi  \exp \left(\zeta + \displaystyle\frac {\mu^\prime(s_{(t+1)j}) - (\mu_{bp}(s_{(t+1)j}) + b_{u})} 
    {b_{u} - b_{l}} \right)^{-1}, \quad \\
    \quad if \ \mu^\prime(s_{(t+1)j}) > \mu_{bp}(s_{(t+1)j}) + b_{u} \\ 
    \\ \varpi  \exp \left(\zeta + \displaystyle\frac{(\mu_{bp}(s_{(t+1)j}) + b_{l}) - \mu^\prime(s_{(t+1)j})} 
    {b_{u} - b_{l}} \right)^{-1},  \\
    \quad if \ \mu^\prime(s_{(t+1)j}) < \mu_{bp}(s_{(t+1)j}) + b_{l} 
  \end{array}
  \right.
\end{align}

\subsection{Method Extension}
Building on EnhancedRL, we further implement user interest augmentation and item representation augmentation via stream clustering and memory networks 
presented in \cite{ref42}, achieving obvious improvements. We also test replacing the clustering algorithm used in \cite{ref42} with alternative clustering algorithms, 
such as VQ-VAE \cite{ref48}, RQ-VAE \cite{ref49} and FSQ \cite{ref50}.
To ensure a fair comparison between EnhancedRL and existing RL-MTF algorithms, the version presented in this paper does not include these modifications.

\subsection{Recommender System with RL-MTF}
\label{sec:deploy}
\begin{figure}[hbtp!]
  \centering
  \includegraphics[width=0.95\linewidth]{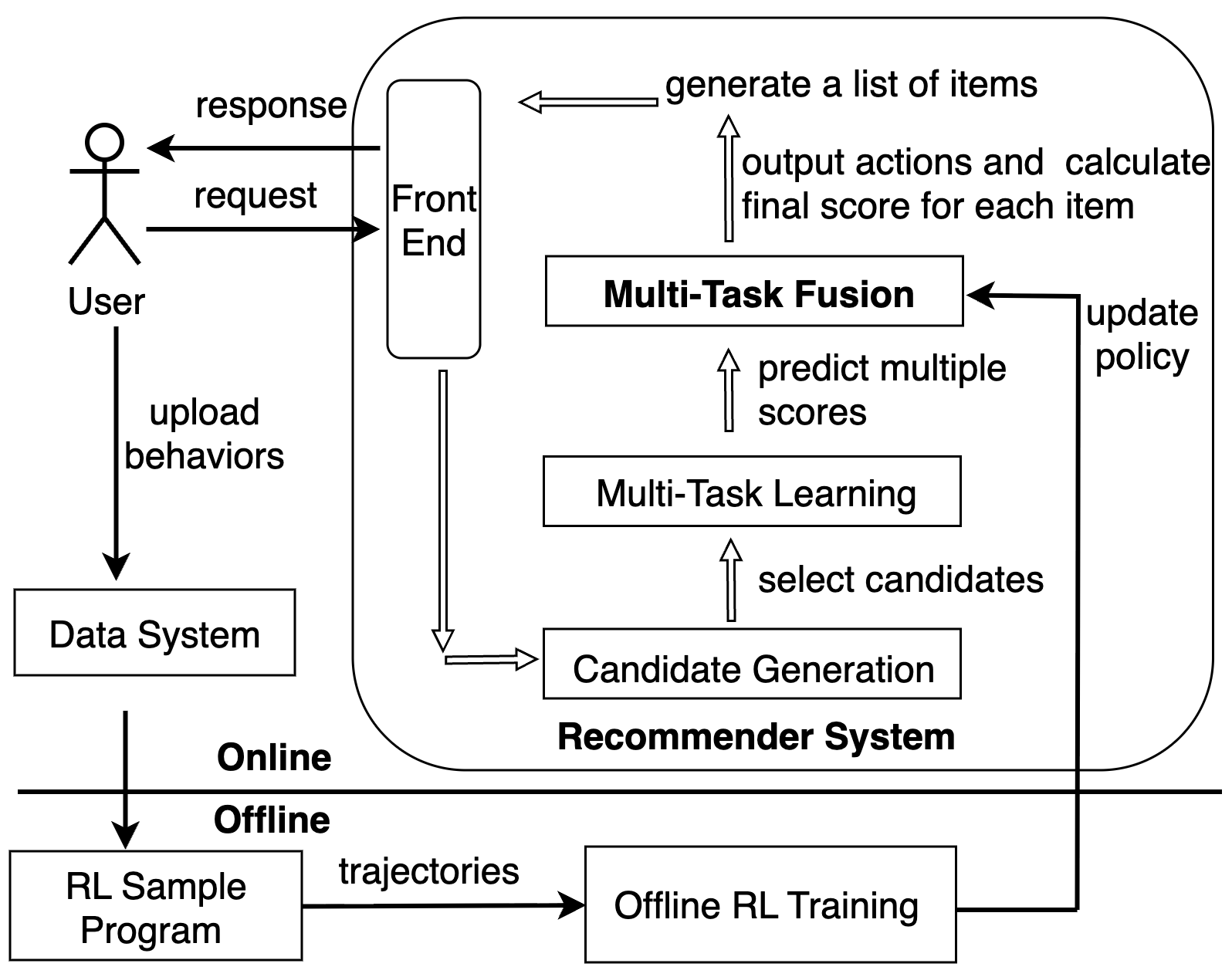}
  \caption{RL-MTF framework in recommender system.}
  \label{fig:implement}
\end{figure}

We implement EnhancedRL in an industrial RS, as shown in Figure \ref{fig:implement}. 
The RL-MTF framework consists of two components: offline model training and online model serving.
The offline model training component is responsible for preprocessing exploration data and training RL model.
The online model serving component generates personalized action for each user-item pair.
Additionally, the online model serving component takes charge of online exploration to collect training data.

%%%%%%%%%%%%%%%%%%%%%%%%%%%%%%%%%%%%%%%%%%%%%%%%%%%%%%%%%%%%%%%%%%%%%%%%%%%%%%%%%%%%%%
%%%%%%%%%%%%%%%%%%%%%%%%%%%%%%       Experiments     %%%%%%%%%%%%%%%%%%%%%%%%%%%%%%%%%
%%%%%%%%%%%%%%%%%%%%%%%%%%%%%%%%%%%%%%%%%%%%%%%%%%%%%%%%%%%%%%%%%%%%%%%%%%%%%%%%%%%%%%
\section{Experiments}
\label{sec:experiments}
EnhancedRL is proposed after UnifiedRL, which has already been deployed in our RS as well as in several other industrial RSs.
To avoid redundant work, in the first part of experiments, we directly reference the results from \cite{ref41}, which compared UnifiedRL with previous MTF algorithms.
Then, in the second part of experiments, we conduct a careful comparison between EnhancedRL and UnifiedRL.

\subsection{Evaluation Metrics}
\subsubsection{Offline Evaluation Metrics}
Following \cite{ref46}, we utilize Normalized Capped Importance Sampling (NCIS) \cite{ref47} to evaluate model performance. 
The cumulative reward across all test user trajectories serves as the evaluation metric. 
Considering that NCIS requires the critic of each RL model to estimate the cumulative reward, that may introduce bias among different critics and 
can only be used for comparison between RL models. 
Since the goal of MTF is to generate a final ranking sequence, similar evaluation metrics can be used as well. 
Inspired by this, to compare various MTF methods, not only RL-MTF methods but also other types of MTF methods, 
\cite{ref41} propose a new evaluation metric called MTF-GAUC build on GAUC \cite{ref34}, as shown in Eq. \ref{eq:mtf_guac}. 
The advantage of MTF-GAUC is that it's simple and better matched to RS. 
Both of them are used as the offline evaluation metrics in this paper.

\begin{equation}
\label{eq:mtf_guac}
  MTF\!\!-\!\!GAUC = \frac{\sum_{(u)}w_{(u)} * Weighted\_AUC_{(u)}}{\sum_{(u)}w_{(u)}} \,
\end{equation}

\subsubsection{Online A/B Testing Metrics}
We evaluate each model using user valid consumption and user duration time, 
which are the two most important online metrics in our recommendation scenario.

\textbf{User Valid Consumption (UVC):} The average total valid consumption per user in a day. 
A valid consumption is defined as a user watching a video for more than $10$ seconds.

\textbf{User Duration Time (UDT):} The average total watching time per user within a day.

\subsection{Experiments One}
\label{experiment1}

\subsubsection{Compared Methods}
UnifiedRL is compared to ES and other previously proposed RL-MTF methods.

• \textbf{ES} \cite{ref20, ref21, ref22, ref23} uses user features as input to generate personalized fusion weights, which is used as the baseline for comparison.

• \textbf{BatchRL-MTF} \cite{ref15} is proposed for MTF in RSs and generates actions based on BCQ \cite{ref28}, which is widely used at Tencent. 
The other RL-MTF methods are implemented based on the framework proposed by BatchRL-MTF.

• \textbf{DDPG} (Deep Deterministic Policy Gradient) \cite{ref27, ref35} is a classical off-policy actor-critic algorithm that can 
learn policies in high-dimensional, continuous action spaces.

• \textbf{CQL+SAC} (Conservative Q-Learning with Soft Actor-Critic) \cite{ref30, ref39} learns a conservative, lower-bound Q function by 
regularizing the Q value of OOD action-state pairs.

• \textbf{IQL} (Implicit Q-learning) \cite{ref31} does not require evaluating actions outside of the dataset, 
but still enables the learned policy to substantially improve over the best behavior in the data through generalization.

• \textbf{UnifiedRL} \cite{ref41} seamlessly integrates offline model with its custom exploration policy and is trained through frequent iterations of online exploration 
and offline training.

\subsubsection{Offline Evaluation}
To compare the performance of the above MTF methods, their models are trained separately and evaluated using both NCIS and MTF-GAUC, 
as shown in Table \ref{table:offline_auc1} and Table \ref{table:offline_auc2}. ES is taken as baseline and is not an RL model, 
therefore NCIS is not used to evaluate ES. The results of these two evaluation approaches are consistent.
\begin{table}[hbtp!]
    \caption{The cumulative reward of the compared RL-MTF methods.}
    \label{table:offline_auc1}
    \footnotesize
    \renewcommand\arraystretch{1.1}{
    \begin{minipage}{\columnwidth}
    \begin{center}
    \resizebox{\linewidth}{!}{
    \small{
    \begin{tabular}{ c|c }
    \toprule
    \textbf{ \qquad Compared Methods \qquad  } & \textbf{\quad \ \ Cumulative Reward \ \ \quad } \\ \hline
    DDPG & 51.62 \\ 
    CQL+SAC & 51.85 \\ 
    BatchRL-MTF & 52.09 \\ 
    IQL & 52.39 \\ 
    \textbf{UnifiedRL} & \textbf{\ 53.96} \\
    \bottomrule
    \end{tabular}
    }}
    \end{center}
    \end{minipage}
    }
\end{table}

\begin{table}[hbtp!]
    \caption{The MTF-GAUC of the compared methods on the same test dataset.}
    \label{table:offline_auc2}
    \footnotesize
    \renewcommand\arraystretch{1.1}{
    \begin{minipage}{\columnwidth}
    \begin{center}
    \resizebox{\linewidth}{!}{
    \small{
    \begin{tabular}{ c|c }
    \toprule
    \textbf{\qquad \ \  Compared Methods \ \  \qquad } & \textbf{\qquad \ \  MTF-GAUC \  \  \qquad } \\ \hline
    ES & 0.7836 \\ 
    DDPG & 0.7881 \\ 
    CQL+SAC & 0.7892 \\ 
    BatchRL-MTF & 0.7894 \\ 
    IQL & 0.7906 \\ 
    \textbf{UnifiedRL} & \textbf{0.7953} \\ 
    \bottomrule
    \end{tabular}
    }}
    \end{center}
    \end{minipage}
    }
  \end{table}
In offline evaluation, the MTF-GAUC of DDPG is higher than that of ES model, 
due to its more powerful performance and consideration of long-term rewards, as shown in Table \ref{table:offline_auc2}.
As previously mentioned, to avoid OOD, existing offline RL algorithms impose overly strict constraints, 
which significantly impairs their performance. This is because these algorithms are limited to training on a fixed dataset and 
are unaware of the exploration policy that generated the data. 
Therefore, they can only avoid OOD problem through excessively strict constraints.
In UnifiedRL, a custom exploration policy is designed and during offline model training, 
the upper and lower bounds of the exploration distribution can be directly obtained, 
which is leveraged to simplify the overly strict constraints and 
integrate the offline model algorithm with its efficient exploration policy. 
In this way, UnifiedRL significantly improves its model performance and outperforms other existing MTF methods.

\subsubsection{Online Evaluation}
\begin{table}[hbtp!]
    \caption{The online results of the compared methods in an industrial RS}
    \label{table:online_res}
    \footnotesize
    \renewcommand\arraystretch{1.1}{
    \begin{minipage}{\columnwidth}
    \begin{center}
    \resizebox{\linewidth}{!}{
    \small{
    \begin{tabular}{ c|c|c }
    \toprule
    \textbf{\quad Compared Methods \quad} & \textbf{\quad \ \ \ UVC \ \ \  \quad} & \textbf{\quad \ \ \ UDT \ \ \  \quad} \\ \hline
    ES & * & * \\ 
    DDPG & +1.39\% & +0.81\%  \\ 
    CQL+SAC & +1.62\% & +0.95\% \\ 
    BatchRL-MTF & +1.79\% & +0.98\% \\ 
    IQL & +2.09\% & +1.15\% \\ 
    \textbf{\qquad UnifiedRL \qquad } & \textbf{ +4.64\%  } & \textbf{ +1.74\% } \\
    \bottomrule
    \end{tabular}
    }}
    \end{center}
    \end{minipage}
    }
\end{table}

The results of online experiments in large-scale RSs are the ultimate criterion for evaluating the effectiveness of a method. 
Therefore, all the compared models are deployed in an industrial RS for one week to conduct A/B tests. 
ES is used as the baseline to demonstrate the improvements of the other models. 
The results of the online experiments are shown in Table \ref{table:online_res}, and all improvements are statistically significant with 
$p$-values less than $0.05$.

DDPG increases UVC by $+1.39\%$ and UDT by $+0.81\%$ compared to ES, 
benefiting from its much stronger model performance and consideration of long-term rewards. 
CQL+SAC outperforms DDPG, and BatchRL-MTF slightly outperforms CQL+SAC. 
IQL surpasses BatchRL-MTF, achieving a $+2.09\%$ increase in UVC and a $+1.15\%$ 
increase in UDT compared to ES. 
UnifiedRL significantly outperforms all other methods, with a $+4.64\%$ increase in UVC and 
a $+1.74\%$ increase in UDT compared to the baseline. 
Moreover, UnifiedRL has also been applied to several other large-scale RSs, achieving significant improvements.

\subsection{Experiments Two}
\label{sec:experiment2}
In this part, we provide a detailed comparison of EnhancedRL and UnifiedRL.
Since there is no public dataset for MTF, we collect training data from an industrial RS that serves hundreds of millions of users, 
using the exploration approach described in Section \ref{sec:online_exploration}. 
One group of users is used to collect exploration data for UnifiedRL's periodic update, 
which has been deployed in our RS and is used as a baseline. The other group of users is used to collect exploration data for EnhancedRL.
Each group contains about 2 million users that are selected randomly.

\subsubsection{Implementation Details} 
In EnhancedRL, an enhanced-state includes the user's profile features (e.g., age, gender, top k interests, flush num, etc.) and user history behavior sequences,
item features, predictions output by MTL and other context information. 
The action generated by MTF model is a $10$-dimensional vector representing the fusion weights in Eq. \ref{eq:merge_eq}, corresponding to a user-item pair. 
All networks of RL are Multi-Layer Perceptron (MLP) and are optimized based on Adam optimizer \cite{ref43}. 
The set number of critics $q$ is set to $2$ and each set contains $10$ independent critics.
The soft update rate and the delay update step for target networks are $w = 0.08$ and $L = 15$. In addition, the
mini-batch size and training epochs are set to $256$ and $300, 000$ respectively. 
Through hyperparameter search, the hyperparameters of each model is set to their optimal values, which have a significant impact on model performance. 
The details of hyperparameter selection are not listed due to page limitation.

\begin{table}[hbtp!]
    \caption{The cumulative reward of EnhancedRL and UnifiedRL on the same dataset.}
    \label{table:off_enhanced1}
    \footnotesize
    \renewcommand\arraystretch{1.3}{
    \begin{minipage}{\columnwidth}
    \begin{center}
    \resizebox{\linewidth}{!}{
    \small{
    \begin{tabular}{ c|c }
    \toprule
    \textbf{ \qquad Compared Methods \qquad  } & \textbf{\quad   Cumulative Reward   \quad } \\ \hline
    UnifiedRL & 53.98 \\ 
    \textbf{EnhancedRL} & \textbf{\ 56.13} \\
    \bottomrule
    \end{tabular}
    }}
    \end{center}
    \end{minipage}
    }
\end{table}

\begin{table}[hbtp!]
    \caption{The MTF-GAUC of EnhancedRL and UnifiedRL on the same test dataset.}
    \label{table:off_enhanced2}
    \footnotesize
    \renewcommand\arraystretch{1.3}{
    \begin{minipage}{\columnwidth}
    \begin{center}
    \resizebox{\linewidth}{!}{
    \small{
    \begin{tabular}{ c|c }
    \toprule
    \textbf{\qquad \ \ Compared Methods \ \ \qquad } & \textbf{\qquad \   MTF-GAUC \   \qquad } \\ \hline
    UnifiedRL & 0.7954 \\ 
    \textbf{EnhancedRL} & \textbf{0.8037} \\ 
    \bottomrule
    \end{tabular}
    }}
    \end{center}
    \end{minipage}
    }
  \end{table}

\subsubsection{Offline Evaluation}
We train EnhancedRL model and compare its performance with UnifiedRL on the same test dataset using the metrics mentioned earlier. 
The results are shown in Table \ref{table:off_enhanced1} and Table \ref{table:off_enhanced2}.
In offline evaluation, the cumulative reward and MTF-GAUC of EnhancedRL are obviously higher than those of UnifiedRL. 
It demonstrates that EnhancedRL takes advantage of the enhanced states to produce better actions for user-item pairs, which is superior to UnifiedRL generating
only one action for each user request used for all candidates.

\subsubsection{Online Evaluation}
To conduct online A/B test, both models are deployed in a large-scale RS for one week. 
UnifiedRL has been fully deployed in the RS and is taken as the baseline for comparison.
The online results are shown in Table \ref{table:online_res2} and all the improvements have statistical significance with p-value less than $0.05$. 
EnhancedRL significantly outperforms UnifiedRL, increasing $+3.84\%$ UVC and $+0.58\%$ UDT. 
It should be noted that EnhancedRL requires more online resources than UnifiedRL. 
We evaluate the resource requirement for EnhancedRL and it is acceptable for deployment.
\begin{table}[hbtp!]
    \caption{The online results of EnhancedRL and UnifiedRL in an industrial RS}
    \label{table:online_res2}
    \footnotesize
    \renewcommand\arraystretch{1.3}{
    \begin{minipage}{\columnwidth}
    \begin{center}
    \resizebox{\linewidth}{!}{
    \small{
    \begin{tabular}{ c|c|c }
    \toprule
    \textbf{\quad \   Compared Methods  \ \quad} & \textbf{\quad  \ \ UVC \  \  \quad} & \textbf{\quad \  \ UDT  \ \  \quad} \\ \hline
    UnifiedRL & *  & * \\ 
    \textbf{\qquad EnhancedRL \qquad } & \textbf{ +3.84\%  } & \textbf{ +0.58\% } \\
    \bottomrule
    \end{tabular}
    }}
    \end{center}
    \end{minipage}
    }
\end{table}

%%%%%%%%%%%%%%%%%%%%%%%%%%%%%%%%%%%%%%%%%%%%%%%%%%%%%%%%%%%%%%%%%%%%%%%%%%%%%%%%%%%%%%
%%%%%%%%%%%%%%%%%%%%%%%%%%%%%%       Conclusion     %%%%%%%%%%%%%%%%%%%%%%%%%%%%%%%%%%
%%%%%%%%%%%%%%%%%%%%%%%%%%%%%%%%%%%%%%%%%%%%%%%%%%%%%%%%%%%%%%%%%%%%%%%%%%%%%%%%%%%%%%
\section{Conclusion}
\label{conclusion}
In this paper, we first point out the problem that the existing RL-MTF methods so far can only utilize user features as RL state 
to generate action for all candidates but are unable to make use of item features and other valuable features, which leads to suboptimal performance. 
To solve this problem, we propose EnhancedRL. Unlike the modeling pattern of existing RL-MTF methods, 
EnhancedRL first takes user features, item features, and other valuable information jointly as the enhanced state, 
then redefine the actor, the critic and the learning process to take full advantage of the enhanced state to make much better action for each user-item pair to maximize 
cumulative reward within a session. 
EnhancedRL differs significantly from existing RL-MTF solutions as it effectively defines a concept of hierarchical state and proposes 
an innovative learning pattern. We conduct extensive offline and online experiments in an industrial RS. 
The online A/B testing results demonstrate that EnhancedRL outperforms other models significantly.

\section*{GenAI Usage Disclosure}
I know that the ACM's Authorship Policy requires full disclosure of all use of generative AI tools in all stages of the research 
(including the code and data) and the writing. No GenAI tools were used in any stage of the research, nor in the writing.

\balance
% \bibliography{main}

\end{document}